\documentclass[twocolumn,utf8]{jetpl}

\usepackage{cite}

\usepackage{amssymb}
\usepackage{amsmath}
\usepackage{amsfonts}
\usepackage{graphicx}
\usepackage{bm}
\usepackage{xcolor}
\usepackage{subfigure} 
\usepackage{comment}
\usepackage
{hyperref}

\DeclareMathOperator{\sgn}{sgn}

\lat

\begin{document}

\title{Bound states and scattering of magnons on a superconducting vortex \\ in ferromagnet-superconductor heterostructures}

\rtitle{Bound states and scattering of magnons on a superconducting vortex in ferromagnet-superconductor ...}


\author{D.\,S.\, Katkov$^{\, a,b}$, S.\,S.\, Apostoloff$^{\, a,c}$, I.\,S.\, Burmistrov$^{\, a,c}$\thanks{e-mail: burmi@itp.ac.ru.}
}
\address{$^{a}$ \mbox{L. D. Landau Institute for Theoretical Physics, Semenova 1-a, 142432, Chernogolovka, Russia} \\
$^{b}$ Moscow Institute for Physics and Technology, 141700, Moscow, Russia \\
$^{c}$ Laboratory for Condensed Matter Physics, HSE University, 101000, Moscow, Russia
}


\abstract{We study the magnon spectrum in a thin ferromagnetic-superconductor heterostructure in the presence of a single superconducting vortex. We employ the Bogolubov-de Gennes Hamiltonian which describes the magnons in the presence of the stray magnetic field and the non-uniform magnetic texture induced by the vortex. We find that the vortex localizes magnon states approximately in the same way as a charge center produces electron bound states due to screened Coulomb interaction in the two-dimensional electron gas. The number of these localized states is substantially determined by the material parameters of the ferromagnetic film only. We solve the scattering problem for an incident plane spin wave and compute the total and transport cross sections. We demonstrate that the vortex-induced non-uniform magnetic texture in chiral ferromagnetic film produces a skew scattering of magnons. We explore the peculiarities of the quantum scattering problem that correspond to orbiting in the classical limit. }


\maketitle

The interaction between magnetism and superconductivity has attracted the interest of researchers for over nearly half a century \cite{Varma1979}. In recent decades, the focus has shifted towards studying physical phenomena in superconductor-ferromagnet (SF) heterostructures
~\cite{Ryazanov2004,Lyuksyutov2005,Buzdin2005,Bergeret2005,Eschrig2015,Back2020,Gobel2021,Zlotnikov}. Both subsystems in such heterostructures can host topological objects: vortices in a superconducting film and skyrmions in a ferromagnetic film \cite{Bogdanov1989}. Skyrmions and superconducting vortices form bound pairs due to interplay of spin-orbit coupling and proximity effect \cite{Hals2016,Baumard2019} or due to stray fields \cite{Dahir2019,Menezes2019,Dahir2020,Andriyakhina2021,Andriyakhina2022,Andriyakhina2023,Andriyakhina2024}. 
Recently, such stable skyrmion-vortex coexistence has been experimentally realized in [Ir$_1$Fe$_{0.5}$Co$_{0.5}$Pt$_1$]$_{10}$/MgO/Nb sandwich structure \cite{Petrovic2021,Machain2021} and in
[CoFeB/Ir/Ta]$_7$/Nb structure \cite{Xie2023}. Additional support for these studies comes from the fact that skyrmions \cite{Chen2015,Yang2016,Gungordu2018,Mascot2019,Garnier2019,Gungordu2022} and skyrmion-vortex pairs \cite{Rex2019,Rex2020}
predicted to
host Majorana modes which were proposed to be used for platform of scalable topological quantum computations
\cite{Nothhelfer2022,Rokhinson2023}.

For a long time spin waves have been recognized as an interesting and useful tool for studying
magnetism 
\cite{SpinWavesBook}. Recently, significant progress has been made in using spin waves as information carriers at the nanoscale, leading to the emergence of a new field called magnonics \cite{magnonics2015,magnonics2021,Brataas2020}. To control and manipulate spin waves either extrinsic (e.g. graded magnetization \cite{Nikitov2015}) or intrinsic (e.g. domain walls \cite{Braun1994,Wulfhekel2004,Dijken2018,Laliena2022}, skyrmions \cite{Nagaosa2014,Garst2014,Aristov2015}) magnetic textures are used.   

It is well-known \cite{Varma1998,Braude2004} that superconducting order parameter fluctuations affect spin waves in ferromagnetic superconductors \cite{Varma1998,Braude2004}. There is a similar effect in the case of spatially separated superconductivity and magnetism. The magnon spectrum in SF heterostructures is affected by the presence of a superconducting film due to several physical mechanisms: (i) a change in stray fields (alternatively, due to existence of Meissner currents) \cite{Golovchanskiy2018,Golovchanskiy2020,Golovchanskiy2020b,Bauer2022,Silaev2022,Golovchanskiy2023,Borst2023,Kharlan2024}, (ii) the existence of a superconducting vortex lattice \cite{Dobrovolskiy2019,Berakdar2023}, and (iii) the spin-torque effect \cite{Bobkova2022}. 

In this Letter inspired by recent experimental progress \cite{Golovchanskiy2018,Dobrovolskiy2019,Golovchanskiy2020,Borst2023}, we consider a thin SF heterostructure and study the magnon spectrum in a ferromagnetic film in the presence of a single superconducting vortex. In contrast with works \cite{Berakdar2023,Kharlan2024}, we take into account that the homogeneous magnetic texture in a ferromagnetic film is altered \cite{Andriyakhina2023} by 
a Pearl superconducting vortex~\cite{Pearl1964}. Following Ref.~\cite{Garst2014}, we employ the Bogolubov-de Gennes {(BdG)} Hamiltonian which describes the magnons in the presence of the non-uniform magnetic texture and the stray magnetic field, induced by a superconducting vortex. We find that the Pearl vortex localizes magnon states. 
These states resemble electron bound states on charged center due to screened Coulomb interaction in the two-dimensional electron gas \cite{Stern1967}. The number of these localized states ($\approx 10$ for an experimental
setup from
Ref. \cite{Petrovic2021}) is fully determined by the material parameters of the ferromagnetic film only. 

We solve the quantum scattering problem for an incident plane spin wave in the presence of the vortex-induced stray field and non-uniform magnetic texture. We estimate the total and transport scattering cross sections. 
We find that the vortex-induced non-uniform magnetic texture in the chiral ferromagnetic film results in a skew scattering of magnons. 
We demonstrate that the quantum scattering problem has peculiarities since there is orbiting effect in the corresponding classical problem.

\noindent\textsf{\color{blue} Model. --- } We consider a heterostructure consisting of ferromagnetic (top) and superconducting (bottom) films. To neglect the proximity effects, we assume the presence of a thin insulating layer 
(of thickness  
much smaller than the London penetration depth $\lambda_L$) in between films. Motivated by experiments \cite{Petrovic2021,Xie2023}, we consider the superconducting film of thickness 
$d_S{\ll}\lambda_L$.
The superconducting film is assumed to host a Pearl vortex.  We assume also that the ferromagnetic film is thinner than the Pearl length $\lambda{=}\lambda_L^2/d_S$, $d_F{\ll}\lambda$. Then, the free energy of the ferromagnetic film is given in terms of the unit magnetization vector $\bm{m}(\bm{r})$ as
\begin{align}
	&\mathcal{F}[\bm{m}] {=}  d_F \int d^2 \bm{r} \bigl \{ A (\nabla \bm{m})^2 {+} K(1{-} m_z^2)  \notag \\
 &  \quad {} 
 {+} D [m_z \nabla \cdot \bm{m}
 {-} (\bm{m}\cdot \nabla) m_z ] {-} M_s \bm{m}\cdot \bm{B}_{\rm V}|_{z{=}{+}0} \bigr \} .
  \label{eq:MagFe}
\end{align}
Here constants $A{>}0$ and $K{>}0$ stand for the exchange and effective perpendicular anisotropy\footnote{For a thin ferromagnetic film it is possible to include the demagnetizing field contribution into the effective perpendicular anisotropy constant, $K{=}K_0{-}2\pi M_s^2$ \protect\cite{Menezes2019,Andriyakhina2021,Kuznetsov2023}.}, respectively. We added also the Dzyaloshinskii--Moriya interaction (DMI) controlled by
constant $D$.
The $z$ axis is directed perpendicularly to the flims' interfaces. $M_s$ denotes the saturation magnetization. The 
last term in Eq. \eqref{eq:MagFe} describes the Zeeman effect of the magnetic field, generated by the Pearl vortex centered at the origin ($\phi_0{=}h c/2e$ is the flux quantum)~\cite{Pearl1964,AbrikosovBook,Carneiro2000}, 
\begin{eqnarray}
	{\bm B}_{\rm V}  = \phi_0 \sgn(z) \nabla 
	\int \frac{d^2\bm{q}}{(2\pi)^2} \frac{e^{-q |z| +i \bm{q}\bm{r}}}{q(1+2q\lambda)} .
	\label{eq:vortex:B}
\end{eqnarray} 
We note that the free energy~\eqref{eq:MagFe} vanishes for the ferromagnetic state, $m_z{=}1$, in the absence of the Pearl vortex, ${\bm B}_{\rm V}{=}0$. 

In the limit of thin films it is convenient to express the magnetic field in the ferromagnetic film as ${{\bm B}_V|_{z=+0}{=}  
{-}(\phi_0/4\pi l_{w}\lambda) [b_r(r)\bm{e}_r{+}b_z(r){\bm e}_z]}$, where $b_r(r)$ and $b_z(r)$ are the dimensionless functions depending on the radial distance $r$ from the vortex center only,
 $\bm{e}_r$, $\bm{e}_\varphi$, and $\bm{e}_z$ are orthogonal unit vectors in polar coordinates with origin at the Pearl vortex, and $l_w{=}\sqrt{A/K}$ is the length scale of a domain wall width. The functions $b_{r,z}(r)$ are expressed in terms of Bessel functions of the second kind ($Y_\alpha$) and Struve functions ($H_\alpha$),
\begin{equation}
\begin{split}
 b_r &  = - \frac{\pi l_w}{4\lambda}\Bigl [Y_1\left (\frac{r}{2\lambda}\right ) + H_{-1}\left (\frac{r}{2\lambda}\right )\Bigr ] ,  \\
 b_z & = \frac{l_w}{r}\Bigr \{ 1 +\frac{\pi r}{4\lambda} \Bigl [Y_0\left (\frac{r}{2\lambda}\right ) - H_{0}\left (\frac{r}{2\lambda}\right )\Bigr ]  \Bigr \} .
\end{split}
\label{eq:mag:field:exact}
\end{equation}
They have the 
following 
simpler asymptotic expressions: $b_r{\simeq}b_z{\simeq} l_w/r$ for $r{\ll}\lambda$, and 
$b_z{\simeq} 2\lambda b_r/r{\simeq}4\lambda^2l_w/r^3$ for $r{\gg}\lambda$. Below, instead of the exact expressions \eqref{eq:mag:field:exact}, we will use the approximate expressions, cf. Ref. \cite{tanguy2001}, which captures correctly not only asymptotics but also the body of the functions: 
\begin{gather}
b_z \simeq \frac{l_w}{r[1+r/(2\lambda)]^2}, \quad b_r \simeq \frac{l_w}{r[1+r/(2\lambda)]} .
\label{eq:Tanguy}
\end{gather}

\noindent\textsf{\color{blue} Magnetic state perturbed by the Pearl vortex. --- } Due to the absence of azimuthal component of the vortex-induced magnetic field, we seek the solution for the magnetization in the ferromagnetic film in the following form,
$\bm{m} {=} \bm{e}_r \sin \theta (r)  {+} \bm{e}_z \cos \theta (r)$.
Minimizing $\mathcal{F}[\bm{m}]$ in Eq. \eqref{eq:MagFe} with respect to the 
magnetization
angle~$\theta(r)$, we find the Euler-Lagrange equation~\cite{Andriyakhina2023},
\begin{gather}
	\frac{l_{w}^{2}}{r} \partial_r \big[r \partial_r \theta(r)\big]
	-\frac{(l_{w}^{2}+r^2)}{2 r^2}\sin2 \theta(r)    
	+ 2\epsilon \frac{ \sin^2 \theta(r)}{r/l_{w}} 
	\notag\\
	+\gamma  [b_z(r)\sin\theta(r)-b_{r}(r)\cos\theta(r)] = 0 ,
	\label{eq:ELE_theta_coax}
\end{gather}
which is supplemented by the boundary conditions $\theta(0){=}\theta(\infty){=}0$.
The DMI strength is controlled by the dimensionless parameter ${\epsilon{=}D/2\sqrt{AK}}$. The effect of the Pearl vortex is described by dimensionless parameter 
${\gamma{=}(l_{w}/\lambda)(M_s\phi_0/8\pi A)}$. Having in mind that $\gamma{\sim}0.1$ in experiments \cite{Petrovic2021}, we will focus on the case $\gamma{\ll} 1$. 

In the regime $\gamma{\ll}1$, the disturbance of the homogeneous ferromagnetic state, corresponding to $\theta{=}0$, is weak, $\theta{=}\theta_{\gamma}{\ll}1$. Therefore, we can linearize Eq.~\eqref{eq:ELE_theta_coax} and, then find 
\begin{align}
 \theta_\gamma(r) =
 \frac{\pi \gamma  \lambda l_w}{4 \lambda ^2+l_w^2}
 \Big[&
 I_1\Big(\frac{r}{l_w}\Big)
 -{L}_{-1}\Big(\frac{r}{l_w}\Big)
  +\frac{4 \lambda  }{\pi l_w}K_1\Big(\frac{r}{l_w}\Big)
 \notag \\   & 
 {}+ Y_1\Big (\frac{r}{2\lambda}\Big )
 +{H}_{-1}\Big(\frac{r}{2\lambda}\Big)
 \Big] ,
 \label{eq:theta:g:0}
 \end{align}
where $I_\alpha$ and $K_\alpha$ denote the modified Bessel functions of the first and second kind, respectively, and $L_\alpha$ stands for the modified Struve function. The exact solution \eqref{eq:theta:g:0} can be simplifed as 
 \begin{equation}
\theta_\gamma(r)=
 -\gamma\begin{cases} 
 [r/(2 l_w)]\ln (l_w/r), & \quad r\ll l_w  ,\\
 l_w/r, & \quad l_w\ll r\ll \lambda, \\
 2\lambda l_w/r^2, & \quad r\gg\lambda .
\end{cases} 
\label{eq:theta:g}
\end{equation}

\noindent\textsf{\color{blue} Magnons. --- } Magnons are encoded in small deviations $\delta\bm{m}{=}\bm{m}{-}\bm{m}_\gamma$ of magnetization~$\bm{m}$ from the stationary state
$\bm{m}_\gamma$ 
determined by $\theta_\gamma$. In order to describe $\delta\bm{m}$, we follow the approach developed in Ref. \cite{Garst2014}. Let us introduce the local orthonormal basis
$\bm{e}_1{=}\bm{e}_\varphi$, $\bm{e}_2{=}[\bm{m}_\gamma{\times}\bm{e}_\varphi]$, and $\bm{e}_3{=}\bm{m}_\gamma$,
and parametrize the magnetization vector at arbitrary spatial point $\bm{r}$ as
\begin{gather}
\bm{m}=\bm{e}_3 \sqrt{1-2|\psi|^2}+\bm{e}_+\psi+\bm{e}_-\psi^* ,
\label{ansatz}
\end{gather} 
where $\bm{e}_\pm{=}{(\bm{e}_1{\pm} i\bm{e}_2)}/{\sqrt{2}}$ and $\psi(\bm{r})$ is a complex function. Substituting the parametrization~\eqref{ansatz} into the free energy~\eqref{eq:MagFe}, and expanding to the second order in $\psi$, one obtains the spin wave energy. Restoring the dynamical part of the problem, we find the following Lagrangian in the dimensionless imaginary time \cite{Garst2014}
\begin{gather}
\mathcal{L}^{(2)}=\frac{d_F K}{2}(\Psi^\dag\sigma^z \partial_t \Psi+ \Psi^\dag\hat H\Psi )
\label{eq:LL}
\end{gather} 
which governs magnon dynamics. Here we introduce a spinor $\Psi{=}(\psi,\psi^*)^T$, standard Pauli matrices $\sigma^{x,y,z}$,  and the effective Hamiltonian of the BdG type:    
\begin{equation}
\hat H=1+\frac{l_w^2}{r^2}-l_w^2\Delta+2i\sigma^z \frac{l_w^2}{r^2} \partial_\varphi + \hat V , \label{eq:H}
\end{equation}
where $\hat V{=}V_0{+}\sigma^x V_x{+}2i\sigma^z V_z  \partial_\varphi$ is the $2{\times}2$ matrix potential, which components depend on $b_{r,z}$, $\theta_\gamma$, and $r$,
\begin{eqnarray}
V_0  &=& {}
- \gamma({b_z}\cos \theta_\gamma 
+{b_r}\sin \theta_\gamma) 
-\dfrac{3}{2}\dfrac{l_w^2+r^2}{r^2} \sin^2 \theta_\gamma 
\notag \\
&& {}
-\dfrac{l_w^2}{2}(\partial_r\theta_\gamma)^2
-\epsilon\Big[
    \dfrac{3 \sin(2\theta_\gamma)}{2r/l_w}
    + l_w\partial_r \theta_\gamma\Big],
\notag  \\
V_x &{=}& 
\dfrac{l_w^2{+}r^2}{2r^2} \sin^2 \theta_\gamma
{-}\frac{l_w^2}{2}(\partial_r \theta_\gamma)^2
{+}\epsilon\Big[
    \dfrac{\sin(2\theta_\gamma)}{2r/l_w} 
    {-}l_w\partial_r\theta_\gamma\Big], 
\notag \\
V_z &=& {}
-2\dfrac{l_w^2}{r^2}\sin^2(\theta_\gamma/2) 
-\epsilon\dfrac{l_w}{r}\sin\theta_\gamma. 
\label{eq:U0}
\end{eqnarray} 
    
We note that the Hamiltonian~\eqref{eq:H} has the particle-hole symmetry: $\hat H{=}\sigma^x K \hat H \sigma^x K$ where $K$ denotes the complex conjugation. Then, it is convenient to seek the solution of the dynamical equation corresponding to the Lagrangian~\eqref{eq:LL} in the following form: $\Psi {=} e^{{-}t E} \tilde{\Psi} {+} e^{t E} \sigma^x K \tilde{\Psi}$, where $\tilde{\Psi}$ satisfies the BdG-type equation: 
    \begin{equation}
    \hat H \tilde{\Psi}{=}E\sigma^z\tilde{\Psi} . 
    \label{eq:BdG:eq}
    \end{equation}
If $\tilde{\Psi}_E$ is a solution of Eq.~\eqref{eq:BdG:eq} with the energy $E$, then the state $\tilde{\Psi}_{-E}{=}\sigma^x K \tilde{\Psi}_E$ is the solution of the same equation with the energy ${-}E$.

In the absence of the Pearl vortex, $b_{r,z}{=}0$, the homogeneous ferromagnetic state corresponds to $\theta{=}0$, such that the Hamiltonian \eqref{eq:H} transforms to 
\begin{equation}
\hat H_0=1+ l_w^2\left[{}-\partial_r^2-\frac{1}{r}\partial_r+\frac{1}{r^2}(-i\partial_\varphi-\sigma^z)^2\right] .    \label{eq:BdG:eq:H0}
\end{equation}
Since $\hat H_0$ is just the Laplacian operator with shifted angular momentum, 
the corresponding  eigen energies and eigen functions can be easily found
\begin{gather}
E_{k,\pm}=\pm(1+k^2l_w^2), \quad  \tilde{\Psi}_{\bm{k},m,\pm}=\tilde{\Psi}_{\pm} J_{m_\pm}(kr) e^{i m \varphi}, \notag \\
\tilde{\Psi}_{+}=(1,0)^T  ,
\quad 
\tilde{\Psi}_{-}=(0,1)^T  ,
\quad
m_\pm=m\mp1, 
\label{eq:eigen:H0}
\end{gather}
where $J_\alpha$ denotes Bessel function of the first kind.

In the regime $\gamma{\ll}1$, it is enough to expand the potential $\hat{V}$ upto the first order in $\theta_\gamma$ keeping terms proportional to $\gamma$ only. Then we find
\begin{gather}
V_0\simeq -\gamma b_z -\epsilon\Big[ \frac{3l_w  \theta_\gamma}{r} +l_w \partial_r \theta_\gamma\Big], \notag \\
V_x\simeq \epsilon\Big[\frac{l_w \theta_\gamma}{r} -l_w \partial_r \theta_\gamma\Big],
\quad 
V_z\simeq -  \epsilon\frac{l_w  \theta_\gamma}{r}.
\label{eq:Vj:1st}
\end{gather}

Interestingly, at $r{\gg}l_w$ the potential $\hat V$ is dominated by the term proportional to $b_z$: $\hat V{\simeq} {-}\gamma b_z$. Namely, using Eqs. \eqref{eq:Tanguy} and \eqref{eq:theta:g}, one finds that at $l_w{\ll}r{\ll} \lambda$ the terms proportional to $\theta_\gamma$ are small in parameter $l_w/r{\ll}1$, while at $r{\gg}\lambda$ these terms are small in parameter $l_w/\lambda{\ll}1$. Only at $r{\sim} l_w$ one should account them as comparable to the leading approximation $\gamma b_z$. Therefore, we start our research from the approximate Hamiltonian,
\begin{gather}
    \hat H\simeq1+\frac{l_w^2}{r^2}-l_w^2\Delta+2i  \sigma_z  \frac{l_w^2}{r^2} \partial_\varphi -\gamma b_z,
    \label{eq:H:approx:0}
\end{gather}
and then extend the results to the Hamiltonian~\eqref{eq:H} with $\hat{V}$ from Eq.~\eqref{eq:Vj:1st}, calculating the required corrections due to terms proportional to $\theta_\gamma$. 

Note that such type of Hamiltonian as in Eq.~\eqref{eq:H:approx:0} with exact \eqref{eq:mag:field:exact} or approximate \eqref{eq:Tanguy} expressions for $b_z$ were studied extensively in the context of electron bound states on charge centers due to screened Coulomb interaction in two-dimensional electron systems \cite{Stern1967,Portnoi1997,tanguy2001,Portnoi2005,Makowski2011a,Makowski2011b}. In what follows we will use the approximate expression \eqref{eq:Tanguy} for $b_z$.

\noindent\textsf{\color{blue} Localized magnon states. --- } 
The effective screened Coulomb potential, $\hat V{\simeq} {-}\gamma b_z$, localizes the magnon states.
At $r{\ll}\lambda$ Hamiltonian~\eqref{eq:H:approx:0} fits the two-dimensional hydrogen, $\hat V{\simeq} {-}\gamma l_w/r$, problem for a particle with mass $1/(2l_w^2)$ and charge $\sqrt{\gamma l_w}$. The energies and eigen functions of such localized states are given as~\cite{Karnakov} 
\begin{eqnarray}
        &&\tilde{\Psi}_{n,m,\pm}{=}\tilde{\Psi}_\pm e^{i m\varphi{-}\frac{2r}{n a_\gamma}}\left(\frac{r}{n a_\gamma}\right)^{|m_{\pm}|}L_{n_r}^{2|m_{\pm}|}
        \left(\frac{4r}{n a_\gamma}\right), \notag \\
        &&E_{n,m,\pm}{=}{\pm}\Big(1{-}\frac{\gamma^2}{n^2}\Big),
        \, n{=}2n_r{+}2|m_{\pm}|{+}1,
        \, n_r{=}0,1,\dots
        \label{hydro_sol}
\end{eqnarray}
Here $L_n^a(x)$ stands for the Sonine-Laguerre polynomials, $a_\gamma{=}2l_w/\gamma$ is the effective Bohr radius,
and $\tilde{\Psi}_\pm$ and $m_\pm$ are defined in Eq.~\eqref{eq:eigen:H0}.

The leading order approximation of the energy of the localized states within Hamiltonian~\eqref{eq:H:approx:0} is independent of DMI. In order to determine dependence on DMI parameter $\epsilon$, it is necessary to take into account the terms proportional to $\theta_\gamma$ in Eq. \eqref{eq:Vj:1st}. The correspondent corrections appears to be small, since states~\eqref{hydro_sol} are localized at the effective Bohr radius $a_\gamma{\gg} l_w$. Then we can apply the perturbation theory to find them,
\begin{equation}
\delta E_{n,m,\pm}^{\rm (DMI)}\simeq \mp \frac{4\gamma^3\epsilon}{n^3}\sgn(2m_\pm-1).   
\label{eq:corr:DMI}
\end{equation}
Interestingly, the DMI-induced contribution does not fully break degeneracy of the spectrum with respect to angular momentum projection $m$.

We note that the effective Bohr radius~$a_\gamma$ is smaller than Pearl length~$\lambda$ for the relevant experimental setups. Indeed, the relation between these lengths is determined by the material parameters of the ferromagnetic film only, $\zeta{\equiv}2\lambda/a_\gamma{=}{\gamma\lambda}/{l_w}{=}M_s\phi_0/(8\pi A)$, e.g. $\zeta{\approx}8.3$ for the SF heterostructure of Ref.~\cite{Petrovic2021}. Therefore, we will assume $\zeta{\gg}1$ and, equivalently, $a_\gamma{\ll}\lambda$ in what follows.

The characteristic scale of hydrogen-like states~\eqref{hydro_sol} is proportional to $n^2 a_\gamma$. For $n$ larger than $\sqrt{\zeta}$ the eigen functions of the hydrogen-like states spread beyond $\lambda$ and one cannot use $1/r$ asymptotics of $\hat V$. 
In order to see the effect of $1/r^3$ decay of $b_z$ at large distances $r{\gg}\lambda$ on the energies of localized states we apply WKB approximation. In the case of 2D problem \eqref{eq:H:approx:0}, the Bohr-Sommerfeld quantization with the Langer correction reads
\begin{equation}
{\rm Re}\int\limits_0^\infty \frac{dr}{\pi l_w} \sqrt{\frac{4\gamma l_w\lambda^2 }{r\left(r{+}2\lambda\right)^2}{-}\frac{m_\pm^2
}{r^2/l_w^2}{-}|1{\mp} E_{n,m,\pm}|} =n_r{+}\dfrac{1}{2}.
\label{eq:WKB}
\end{equation}

Then we can calculate the corrections to the energies~\eqref{hydro_sol} of hydrogen-like states at $n^2{\ll}\zeta$ that break degeneracy of the spectrum,
\begin{equation}
    \delta E_{n,m,\pm}^{\rm (WKB)}=\pm (\gamma^2/\zeta)[1+3(4m_\pm^2-3n^2)/(16\zeta)].
    \label{eq:corr:hydro-WKB}
\end{equation}
Corrections \eqref{eq:corr:DMI} and~\eqref{eq:corr:hydro-WKB} are comparable if $\epsilon\gamma\zeta^2{\sim} n^5$.

For eigen states which are very close to the continuum spectrum, $1{-}E_{n,m,\pm}{\ll}\gamma^2/\zeta$, and 
have not small angular momentum, $1{\ll} m_\pm^2{\sim}\zeta$, we find
\begin{gather}
E_{n,m,\pm}{\simeq} \pm\Big[1{-}\frac{\gamma^2}{4\zeta}\frac{(2m_\pm^2/\zeta)^{5/2}}{3{-}4m_\pm^2/\zeta}\Big(1{-}\frac{n{+}2|m_\pm|}{2\sqrt{2\zeta}}\Big)\Big].
    \label{eq:WKB:res}
    \end{gather}
Note that the radial quantum number satisfies inequality $0{\leqslant}n_r{\leqslant}{\sqrt{2\zeta}{-}2|m_\pm|}$.
The WKB spectrum \eqref{eq:WKB:res} suggests that only a finite number of localized levels exist. Their number can be estimated in a standard way  to be equal to $\zeta$~\cite{Portnoi1997}.
Physically, finiteness of the number of localized levels is related with the following. For an unscreened Coulomb potential $1/r$ the number of states is infinite but the energy levels are thickening approaching a continuous spectrum. For $\lambda^2/r^3$ potential the number of states is also infinite but the energies are thickening while going away from a continuous spectrum. As a result of competition between $1/r$ and $\lambda^2/r^3$ at $r{\sim}\lambda$, only the finite number of levels, ${\sim}\zeta$, survive.

\noindent\textsf{\color{blue} Plane wave magnon scattering. --- } Let us now discuss the scattering problem for magnons. 
Assuming the incident magnons with the energy $E_{k,\pm}{=}{\pm}(1{+}k^2l_w^2)$ to be described by a plane wave, and setting $1/\lambda$ and $\epsilon$ to zero, we obtain a scattering problem for 2D Coulomb potential, 
see the simplified Hamiltonian~\eqref{eq:H:approx:0} with $b_z{\simeq}l_w/r$.
In this case the differential scattering cross section is known exactly \cite{Lin1997}:  
\begin{equation}
\frac{d\sigma(\varphi)}{d\varphi}=\frac{\gamma\tanh[{\gamma \pi}/(2 kl_w)]}{4  k^2 l_w\sin^2(\varphi/2)} .  
\label{eq:Coulomb:2D}    
\end{equation}
We note that although the transport cross section is finite, the total cross section diverges due to singularity at $\varphi{=}0$ in expression \eqref{eq:Coulomb:2D}. Therefore, we need to take into account that $b_z$ decays as $1/r^3$ at $r{\gg}\lambda$. Then the differential scattering cross section turns out to be regularized at small angles $|\varphi|{\lesssim}1/( k\lambda)$. Applying Born approximation, we can find 
\begin{equation}
\frac{d\sigma_\pm^{\rm (B)}(\varphi)}{d\varphi} {\simeq} \frac{2\pi \zeta^2/k}{(1{+}2 q_\varphi\lambda 
)^2}
\Big(1{+}\dfrac{2\epsilon q_\varphi l_w}{1{+}q_\varphi^2l_w^2} \Big)
.
 \label{eq:Born}
\end{equation}
Here $q_\varphi{=}2 k|\sin(\varphi/2)|$ is the transferred momentum by a scattered magnon. In addition to Hamiltonian~\eqref{eq:H:approx:0},
we took into account the DMI terms from Eq.~\eqref{eq:Vj:1st}, and assuming them to be small, $\epsilon{\ll}1$.  

Note that in the considered case of large parameter $\zeta$, the Born approximation~\eqref{eq:Born} is valid for relatively fast magnons only, $k\lambda{\gg}\zeta{\gg}1$ or, equivalently, $k l_w{\gg}\gamma$. 

Integrating Eq. \eqref{eq:Born} over $\varphi$ 
we find at $kl_w{\gtrsim}1{\gg}\gamma$ the total and transport cross sections as
\begin{equation}
\begin{split}
	\sigma^{\rm (B)} &\simeq 2\pi \zeta^2/( k^2\lambda)\{1
 +\epsilon [\ln (\lambda/l_w)-1](l_w/\lambda)\},  
	\\  
	\sigma^{\rm (B)}_{\rm tr} &\simeq \pi^2 \zeta^2/(2 k^3\lambda^2) [1-
 2 \epsilon \ln (4 k l_w)/(\pi k l_w) ] ,
 \end{split}
 \label{eq:Born:1}
\end{equation} 
In fact, the expression for $\sigma^{\rm (B)}_{\rm tr}$ does not involve $\lambda$, and thus can be derived directly from Eq. \eqref{eq:Coulomb:2D} 
neglecting DMI. 

In order to compute the scattering cross section in other regions of the parameter $k\lambda$, we employ the WKB approximation for the phase shifts $\delta_m$. Introducing two parameters $g{=}\zeta/[2(k\lambda)^2]$ and 
$\alpha_m{=}|m_\pm|/(2k\lambda)$, we write the phase shift as \cite{Karnakov}
\begin{gather}
    \frac{\delta_m}{2\lambda k}=\int\limits_{x_0}^\infty\left(\sqrt{1{+}\frac{g}{x(1{+}x)^2}{-}\frac{\alpha_m^2}{x^2}}{-}1\right)dx{-}x_0{+}\frac{\pi\alpha_m}{2}.\label{def_alpha}
\end{gather}
Here $x_0{>}0$ is the largest in magnitude turning point and we neglect DMI. We note that Eq. \eqref{eq:BdG:eq:H0} suggests that $\delta_m$ depends on the $m_\pm{=}m{\mp}1$ rather than $m$. However, 
the magnetization $\bm{m}$ has a physical meaning, not $\psi$, cf.~Eq.~\eqref{ansatz}. Transition from $\psi$ to $\bm{m}$ results in an additional factor $\exp(\pm i\varphi)$ that effectively shifts $m_\pm$ to $m$.
We mention that $\delta_m$ is an even function of $m_\pm$ such that the scattering amplitude $f(\varphi){=}\sqrt{2/(\pi k)}\sum_m \exp(im\varphi{+}i\delta_m)\sin \delta_m$ has an even-in-$\varphi$ absolute value.
The WKB approximation for the phase shift is valid for the region $1/\zeta{\ll}k\lambda{\ll}\zeta$.

\begin{figure}[t]
\centerline{\includegraphics[width=0.85\columnwidth]{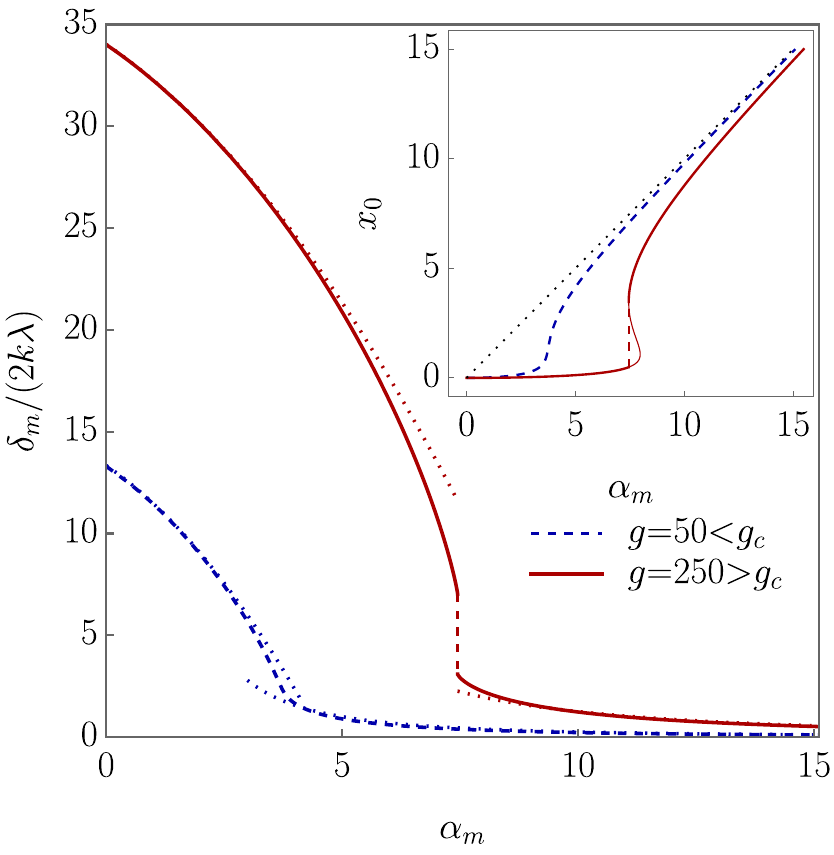}}
\caption{Fig.~\protect\ref{fig:phase:shifts}. Dependence of $\delta_m/(2k\lambda)$ on the parameter $\alpha_m$ for $g{<}g_c$ and $g{>}g_c$ plotted by solid and dashed lines for
the exact integral in Eq.~\eqref{def_alpha} and by dotted lines for its asymptotics from Eq.~\eqref{eq:delta:m:WKB:1}. Inset: Dependence of the turning point $x_0$ on $\alpha_m$ for the same magnitudes of $g$.}
\label{fig:phase:shifts}
\end{figure}

The behavior of the integral in Eq.~\eqref{def_alpha} depends on the magnitude of $g$ in an interesting way. There is a single turning point $x_0$ for $g{<}g_c{=}4(316{+}119\sqrt{7})/27{\approx}93.5$. Consequently, the phase shift is a smooth function of $\alpha_m$. For $g{>}g_c$ and $\alpha$ within a certain interval $(\alpha_g,\alpha_g')$ 
the two additional turning points appear, see inset in Fig.~\ref{fig:phase:shifts}. As a result,
the jump in $\delta_m$ as a function of $\alpha_m$ occurs at $\alpha_g{\geqslant}\alpha_c{=}\sqrt{(37{+}14\sqrt{7})/3}{\approx}4.97$. At $g{\gg}1$ one can estimate 
$\alpha_g{\simeq}\sqrt{3}[(g/2)^{1/3}{-}2/3]$. 
The jump in $\delta_m$ is related to the abrupt appearance of larger root $x_0$
of the expression under the square-root in Eq. \eqref{def_alpha}. 

At $g{\gg}1$ the phase shifts within WKB approximation can be calculated as 
\begin{equation}
 \frac{\delta_m}{2 k\lambda} {=} \begin{cases}
 I_0(g){-}\dfrac{\pi \alpha_m}{2}{-}I_1(g)\alpha_m^2{+}O(\alpha_m^4),
 \!\!\!\!\!\!
 & \, \alpha_m\ll \alpha_g , \\
  g/(2\alpha_m^2), & \, \alpha_m\gg \alpha_g ,  
 \end{cases}   
 \label{eq:delta:m:WKB:1}
 \end{equation}
 where $I_{0,1}(g)$ are estimated as $I_0{\simeq}\pi g^{1/2}{-}2.59 g^{1/3}{+}2/3$ and $I_1{\simeq}1.40g^{-1/3}{-}1.15g^{-2/3}$.
 The exact and asymptotic form of dependence of the phase shift $\delta_m$ on the angular momentum at different magnitudes of $g{\gg}1$ is shown in main panel of Fig.~\ref{fig:phase:shifts}.

\begin{figure}[t]
\centerline{\includegraphics[width=0.95\columnwidth]{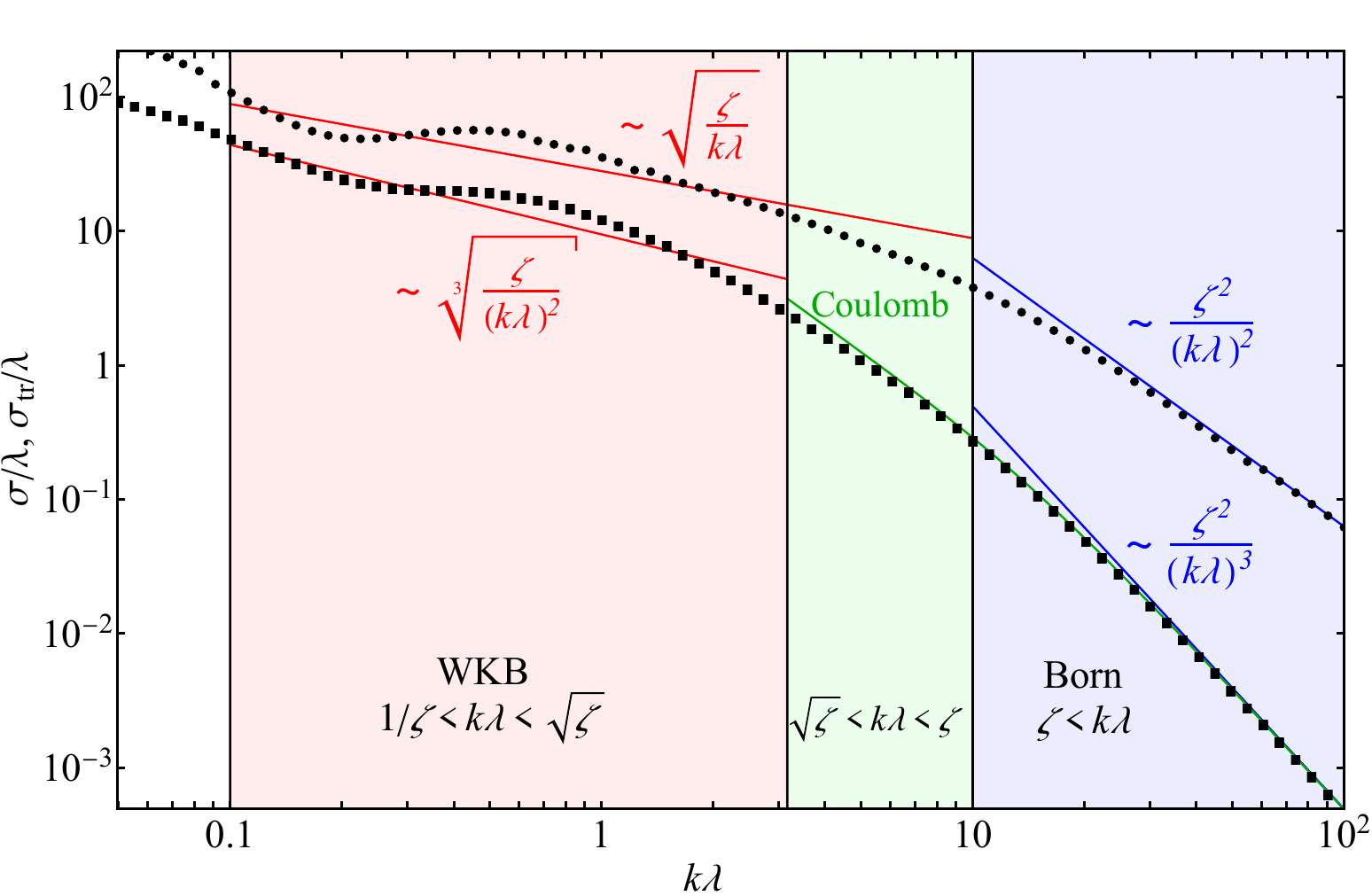}}
\caption{Fig.~\protect\ref{fig:WKB:scattering}. The dependence of the normalized total ($\sigma/\lambda$, the circle marks) and transport ($\sigma_{\rm tr}/\lambda$, the square marks) scattering cross sections on $k\lambda$ for $\zeta{=}10$. The straight lines indicates asymptotic behavior given by Eqs.~\eqref{eq:Born:1} and Eqs.~\eqref{eq:sigma:t:tr:WKB} in Born and WKB approximations, respectively. The curve presents $\sigma_{\rm tr}/\lambda$ calculated from Eq.~\eqref{eq:Coulomb:2D} for 2D Coulomb potential.
}
\label{fig:WKB:scattering}
\end{figure}

\begin{figure}[t]
\centerline{\includegraphics[width=0.9\columnwidth]{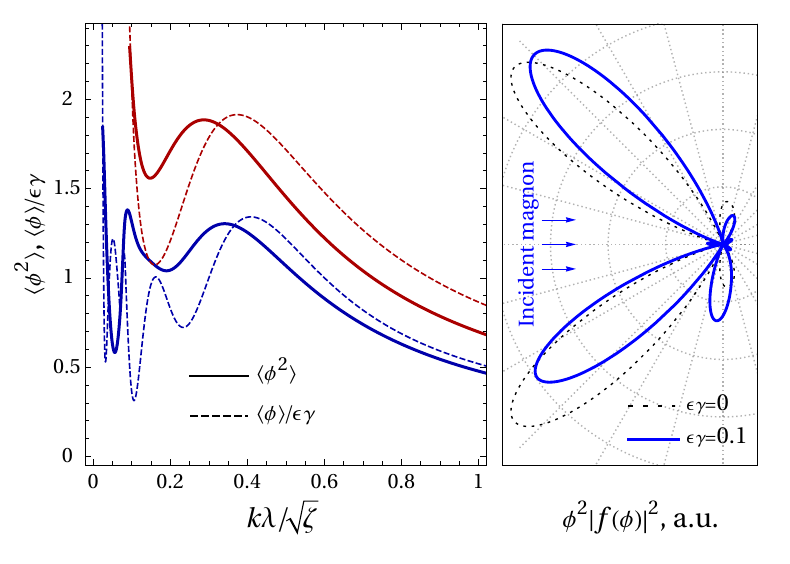}}
\caption{Fig.~\protect\ref{fig:DMI-skew}. \textit{Left panel.} The dependence of $\langle\varphi^2\rangle$ and $\langle\varphi\rangle/\epsilon\gamma$ on normalized momentum $k \lambda/\sqrt{\zeta}$ for $\zeta{=}10$ (upper curves) and $\zeta=30$ (lower curves). \textit{Right panel.} The scattering pattern $d\sigma/d\phi{=}|f(\phi)|^2$ additionally weighed by $\phi^2$ and plotted in polar coordinates for $\zeta{=}30$, $k\lambda{=}0.6$, $\epsilon\gamma{=}0$ (dotted line) and $\epsilon\gamma{=}0.1$ (solid line).
}
\label{fig:DMI-skew}
\end{figure}

The total and transport cross sections can be found using the standard expressions: $\sigma{=}(4/k)\sum_m \sin^2\delta_m$ and $\sigma_{\rm tr}{=}(2/k)\sum_m \sin^2(\delta_m{-}\delta_{m{+}1})$. 
At $g{\gg}1$ the sum is dominated by $m{\sim}(\zeta k\lambda)^\nu{\gg}1$, where $\nu{=}1/2$ and $1/3$ for the total and transport cross sections, respectively. Therefore, using asymptotic expression \eqref{eq:delta:m:WKB:1} for $\alpha_m{\gg}\alpha_g$, we obtain (for $\sqrt{\zeta}{\gg}k\lambda{\gg}1/\zeta$):
\begin{equation}
 \sigma^{\rm (WKB)} \simeq 4 \lambda \sqrt{\frac{2\pi\zeta}{k\lambda}}, \quad \sigma_{\rm tr}^{\rm (WKB)} \sim \lambda \Big [\frac{\zeta}{(k\lambda)^2}\Big]^{1/3}.  
 \label{eq:sigma:t:tr:WKB}
\end{equation}
Note that both total and transport cross sections have contributions oscillating with $k\lambda$, see Figs.~\ref{fig:WKB:scattering} and~\ref{fig:DMI-skew} for details. However, the oscillating part of $\sigma^{\rm (WKB)}$ is parametrically smaller than the monotonous part given by Eq. \eqref{eq:sigma:t:tr:WKB}. In contrast, the oscillating part of $\sigma_{\rm tr}^{\rm (WKB)}$ occurs to be of the same order of magnitude as its monotonous part, Eq. \eqref{eq:sigma:t:tr:WKB}, for a reasonable range of parameters. 

In the WKB region $\sqrt{\zeta}{\ll}k\lambda{\ll}\zeta$ ($g{\ll}1$) we find
\begin{equation}
\frac{\delta_m}{2\lambda k}=\frac{g}{2}\begin{cases}
    \ln[4/(g+2 e \alpha_m )],
    & \alpha_m\ll 1, \\
    1/\alpha_m^2, & \alpha_m\gg 1 .
\end{cases}
\label{eq:delta:m:gg}
\end{equation}
We note that for $\alpha_m{\gg}1$ the dependence of $\delta_m$ on $\alpha_m$ in Eq.~\eqref{eq:delta:m:gg} is the same as in the Born approximation. However, in the considered region of the parameter $k\lambda$ the phase shift $\delta_m$ is not necessarily small. For the total cross section we obtain the same expression as in Eq.~\eqref{eq:sigma:t:tr:WKB}, while for the transport cross section we find the following estimate $\sigma_{\rm tr}^{\rm (WKB)}{\sim}\zeta/(k^2\lambda)$. 

We note that both the total and transport cross sections decay with increase of $k$ for $k\lambda{\gtrsim}\sqrt{\zeta}$. For such momenta the forward scattering is dominant since $\sigma_{\rm tr}/\sigma{\ll}1$ for $k\lambda{\gg}\sqrt{\zeta}$. Figure~\ref{fig:WKB:scattering} shows the dependence of the normalized cross sections on $k\lambda$ calculated both numerically (the circle and square marks) by solving Eq.~\eqref{eq:BdG:eq} with defined angular momentum~$m$ and analytically (solid lines) by different asymptotics in Eqs.~\eqref{eq:Born:1}, \eqref{eq:sigma:t:tr:WKB}, and~\eqref{eq:Coulomb:2D}.

In calculations above
we neglect terms proportional to the DMI parameter $\epsilon$.
Now we take into account the omitted small terms from Eq. \eqref{eq:Vj:1st} to estimate the averaged scattering angle $\langle\varphi\rangle{=}\int_{{-}\pi}^\pi \varphi d\sigma(\varphi)/\sigma$ of magnons to characterize a skew scattering. 
The main skew effect on the scattering amplitude $f(\varphi)$ is produced by~$V_z$, which yields the additional term $\epsilon\gamma m_\pm/[2(k \lambda x)^2(1{+}x)]$ under the square root in the integral in Eq.~\eqref{def_alpha}.
When $\epsilon\gamma{\ll}1$ the average scattering angle $\langle\varphi\rangle$ in main approximation is proportional to $\epsilon\gamma$, then $\langle\varphi\rangle/\epsilon\gamma$ depends only on $k\lambda$ and $\zeta$. 

The left panel of Fig.~\ref{fig:DMI-skew} shows the dependence of the normalized averaged scattering angle $\langle\varphi\rangle/\epsilon\gamma$ (dashed lines) and the averaged squared scattering angle $\langle\varphi^2\rangle$ (solid lines) on normalized momentum $k
\lambda/\sqrt{\zeta}$ for $\zeta{=}10$ (upper curves) and $\zeta{=}30$ (lower curves). Note that the curves show the oscillating behaviour for not large $k\lambda{\lesssim}\sqrt{\zeta}$ and the amplitude of oscillations is comparable to the magnitude of averages itself. These oscillations appears because strong forward scattering is effectively suppressed when one calculates $\langle\varphi\rangle$ and $\langle\varphi^2\rangle$, and the side lobes of skew scattering pattern becomes more important, see the right panel of Fig.~\ref{fig:DMI-skew}. Therefore, the non-uniform magnetic texture induced by the Pearl vortex produces a weak skew scattering in the presence of DMI.

\begin{figure}[t]
\centerline{\includegraphics[width=0.51\columnwidth]{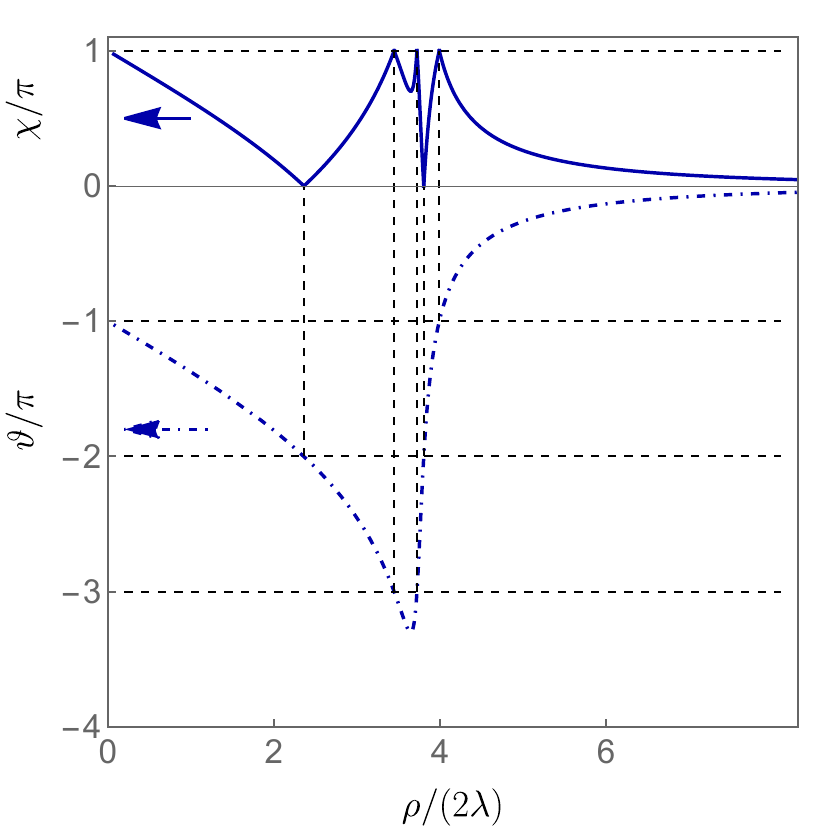}\hspace{-0.41cm}\includegraphics[width=0.484\columnwidth]{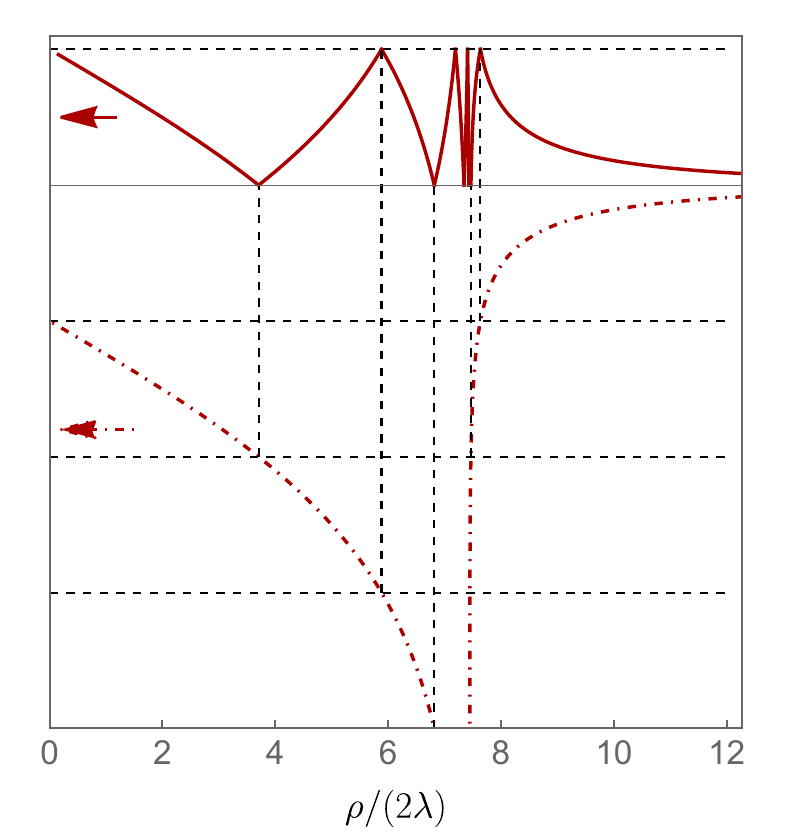}}
\caption{Fig.~\protect\ref{fig:WKB:scatteringAngle}. The angle $\vartheta$ and physical scattering angle $\chi$ as the functions of the impact parameter $\rho/(2\lambda)$ for $g{=}50{<}g_c$ (left panel) and for $g{=}250{>}g_c$ (right panel).}
\label{fig:WKB:scatteringAngle}
\end{figure}

\noindent\textsf{\color{blue} Discussions. --- }  The discontinuity of the phase shift for $g{>}g_c$ at $\alpha_m{=}\alpha_g$, might suggest existence of directions along which there is no scattering wave spreading in WKB approximation. However, this is not the case. Nevertheless, the discontinuity point $\alpha_g$ has a clear physical meaning. As known~\cite{Kotkin}, physical scattering angle $\chi$ as a function of the impact parameter $\rho{=}|m|/k{\simeq}2 \alpha_m\lambda$
can be extracted from the relation ${\pm} \chi{=}2\pi n{-}\vartheta$, where 
$\vartheta {=} 2{d\delta_m}/{dm}{\simeq}(k\lambda)^{-1}{d\delta_m}/{d\alpha_m}$ and $n$ is an integer. Nonzero $n$ implies the phenomenon of orbiting. For $g{<}g_c$ the dependence of $\chi$ and $\vartheta$ on $\rho$ is shown in Fig.~\ref{fig:WKB:scatteringAngle}. For small impact parameter $\rho/(2\lambda){\lesssim}4$, the classical trajectory overturns several (but finite) times around the scattering center. In contrast, for $g{>}g_c$, the number of overturns of classical trajectory around the scattering center tends to infinity as the impact parameter tends to $2\alpha_g\lambda$. In other words, at this impact parameter a falling of the classical particle into the scattering center should occur. We note that $2\alpha_g\lambda$ is always larger than $2\alpha_c\lambda{\approx}10 \lambda$.

Note that in the case of a Pearl anti-vortex, which magnetic field is directed in the opposite direction, the potential \eqref{eq:U0} in Hamiltonian \eqref{eq:H} changes its sign. It results in disappearance of the localized states. Also the scattering problem becomes less interesting, in particular, the effect of orbiting is absent. 

\noindent\textsf{\color{blue} Summary. --- } In this work we considered a thin SF heterostructure and study the magnon spectrum in a ferromagnetic film in the presence of a superconducting Pearl vortex. The vortex induces the stray magnetic field and the non-uniform magnetic texture. It results in the appearance of several localized magnon states, similar to electron bound states, on a charged center, due to the screened Coulomb interaction in a two-dimensional electron gas. The number of these localized states is fully determined by the material parameters of the ferromagnetic film only. 

We solved the scattering problem for an incident plane spin wave. We computed the total and transport cross sections. We explored the peculiarities of the quantum scattering problem that correspond to orbiting in the classical limit. We found that the interplay of the Dzyaloshinskii--Moriya interaction and the vortex-induced non-uniform magnetic texture results in the skew scattering of magnons. 

Our work can be extended in several directions. At first, it would be interesting to consider the scattering of a cylindrical spin wave. Secondly, the localized states of magnons on the bound pair of a Pearl vortex and N\'eel skyrmion is worthwhile to study.

\noindent\textsf{\color{blue} Acknowledgments. --- } We thank Ya. Fominov, A. Kalashnikova, and M. Parfenov for useful discussions. We thank E. Andriyakhina for collaboration on a related project. The work was funded by the Russian Science Foundation under the Grant No. 24-12-00357. 
I.S.B. and D.S.K. acknowledge personal support from the Foundation for the Advancement of Theoretical Physics and Mathematics ``BASIS''.


\begin{thebibliography}{99}

\bibitem{Varma1979} E. I. Blount and C. M. Varma, {\it Electromagnetic effects near the superconductor-to-ferromagnet transition},
Phys. Rev. Lett. {\bf 42}, 1079 (1979).

\bibitem{Ryazanov2004} V. V. Ryazanov, V. A. Oboznov, A. S. Prokofiev,
V. V. Bolginov, and A. K. Feofanov, {\it Superconductor--ferromagnet--
superconductor $\pi$-junctions}, J. Low Temp. Phys. {\bf 136}, 385 (2004).

\bibitem{Lyuksyutov2005} I. F. Lyuksyutov and V. L. Pokrovsky, {\it  Ferromagnet--superconductor
hybrids}, Adv. Phys. {\bf 54}, 67 (2005).

\bibitem{Buzdin2005} A. I. Buzdin, {\it Proximity effects in superconductor--ferromagnet
heterostructures}, Rev. Mod. Phys. {\bf 77}, 935 (2005).

\bibitem{Bergeret2005} F. S. Bergeret, A. F. Volkov, and K. B. Efetov,
{\it Odd triplet superconductivity and related phenomena
in superconductor--ferromagnet structures}, Rev. Mod. Phys. {\bf 77}, 1321 (2005).

\bibitem{Eschrig2015} M. Eschrig, {\it Spin-polarized supercurrents for spintronics:
A review of current progress}, Rep. Prog. Phys. {\bf 78}, 104501 (2015).

\bibitem{Back2020} C. Back, V. Cros, H. Ebert, K. Everschor-Sitte, A. Fert,M. Garst, T. Ma, S. Mankovsky, T. L. Monchesky, M. Mostovoy, N. Nagaosa, S. S. P. Parkin, C. Pffeiderer, N. Reyren, A. Rosch, Y. Taguchi, Y. Tokura, K. von Bergmann, and J. Zang, {\it The 2020 skyrmionics roadmap}, J. Phys. D: Applied Phys. {\bf 53}, 363001 (2020).

\bibitem{Gobel2021} B. G\"obel, I. Mertig, and O. A. Tretiakov, {\it Beyond
skyrmions: Review and perspectives of alternative magnetic
quasiparticles}, Phys. Rep. {\bf 895}, 1 (2021).

\bibitem{Zlotnikov} A. O. Zlotnikov, M. S. Shustin, and A. D. Fedoseev, {\it Aspects of topological superconductivity in 2D systems: Noncollinear magnetism, skyrmions, and higher-order topology}, J. Supercond. Nov. Magn. {\bf 34}, 3053 (2021).

\bibitem{Bogdanov1989} A. N. Bogdanov and D. Yablonskii, {\it Thermodynamically
stable ``vortices'' in magnetically ordered crystals. The
mixed state of magnets}, Sov. Phys. JETP {\bf 68}, 101 (1989).

\bibitem{Hals2016} K. M. D. Hals, M. Schecter, and M. S. Rudner,
{\it Composite topological excitations in ferromagnet--superconductor
heterostructures}, Phys. Rev. Lett. {\bf 117}, 017001 (2016).

\bibitem{Baumard2019} J. Baumard, J. Cayssol, F. S. Bergeret, and A. Buzdin,
{\it Generation of a superconducting vortex via N\'eel
skyrmions}, Phys. Rev. B {\bf 99}, 014511 (2019).


\bibitem{Dahir2019} S. M. Dahir, A. F. Volkov, and I. M. Eremin, {\it Interaction
of skyrmions and Pearl vortices in superconductor -- chiral
ferromagnet heterostructures}, Phys. Rev. Lett.
{\bf 122}, 097001 (2019).

\bibitem{Menezes2019} R. M. Menezes, J. F. S. Neto, C. C. de Souza Silva,
and M. V. Milo\'sevi\'c, {\it Manipulation of magnetic
skyrmions by superconducting vortices in ferromagnet--superconductor
heterostructures}, Phys. Rev. B {\bf 100},
014431 (2019).

\bibitem{Dahir2020} S. M. Dahir, A. F. Volkov, and I. M. Eremin, {\it Meissner
currents induced by topological magnetic textures
in hybrid superconductor/ferromagnet structures}, Phys.
Rev. B {\bf 102}, 014503 (2020).


\bibitem{Andriyakhina2021} E. S. Andriyakhina and I. S. Burmistrov, {\it Interaction of a N\'eel-type skyrmion with a superconducting vortex}, Phys. Rev. B {\bf 103}, 174519 (2021).

\bibitem{Andriyakhina2022} E. S. Andriyakhina, S. Apostoloff, and I. S. Burmistrov,
{\it Repulsion of a N\'eel-type skyrmion from a pearl vortex
in thin ferromagnet–superconductor heterostructures}, JETP Letters {\bf 116}, 825 (2022).

\bibitem{Andriyakhina2023} S. S. Apostoloff, E. S. Andriyakhina, P. A. Vorobyev,
O. A. Tretiakov, and I. S. Burmistrov, {\it Chirality inversion
and radius blowup of a N\'eel-type skyrmion by a Pearl
vortex}, Phys. Rev. B {\bf 107}, L220409 (2023).

\bibitem{Andriyakhina2024} S. S. Apostoloff, E. S. Andriyakhina, and I. S.
Burmistrov, {\it Deformation of a N\'eel-type skyrmion in
a weak inhomogeneous magnetic field: Magnetization
ansatz and interaction with a Pearl vortex}, Phys.
Rev, B {\bf 109}, 104406 (2024).

\bibitem{Petrovic2021} A. P. Petrovi\'c, M. Raju, X. Y. Tee, A. Louat, I. Maggio-Aprile,
R. M. Menezes, M. J. Wyszy\'nski, N. K. Duong, M. Reznikov,
Ch. Renner, M. V. Milosevi\'c, and C. Panagopoulos, {\it Skyrmion-
(Anti)Vortex Coupling in a Chiral Magnet-Superconductor
Heterostructure}, Phys. Rev. Lett. {\bf 126}, 117205 (2021).

\bibitem{Machain2021} P. Machain, {\it Skyrmion-vortex interactions in chiral-
magnet/superconducting hybrid systems}, Nanyang
Technological University (2021).

\bibitem{Xie2023} Y. Xie, A. Qian, B. He, Y. Wu, S. Wang, B. Xu,
G. Yu, X. Han, and X. Qiu, {\it Visualization of skyrmion-
superconducting vortex pairs in a chiral magnet-
superconductor heterostructure}, arXiv:2310.13363  (2023).

\bibitem{Chen2015} W. Chen and A. P. Schnyder, {\it Majorana edge states
in superconductor-noncollinear magnet interfaces}, Phys.
Rev. B {\bf 92}, 214502 (2015).

\bibitem{Yang2016} G. Yang, P. Stano, J. Klinovaja, and D. Loss, {\it Majorana
bound states in magnetic skyrmions}, Phys. Rev. B {\bf 93},
224505 (2016).

\bibitem{Gungordu2018} U. G\"ung\"ord\"u, S. Sandhoefner, and A. A. Kovalev, {\it Stabilization
and control of majorana bound states with
elongated skyrmions}, Phys. Rev. B {\bf 97}, 115136 (2018).

\bibitem{Mascot2019} E. Mascot, S. Cocklin, S. Rachel, and D. K. Morr, {\it Dimensional
tuning of majorana fermions and real space
counting of the Chern number}, Phys. Rev. B {\bf 100}, 184510 (2019).

\bibitem{Garnier2019} M. Garnier, A. Mesaros, and P. Simon, {\it Topological
superconductivity with deformable magnetic skyrmions}, Commun. Phys. {\bf 2}, 126 (2019).

\bibitem{Gungordu2022} U. G\"ung\"ord\"u
and A. A. Kovalev, {\it Majorana bound states with chiral magnetic textures}
J. of Appl. Phys. {\bf 132}, 041101 (2022).

\bibitem{Rex2019} S. Rex, I. V. Gornyi, and A. D. Mirlin, {\it Majorana bound
states in magnetic skyrmions imposed onto a superconductor}, Phys. Rev. B {\bf 100}, 064504 (2019).

\bibitem{Rex2020} S. Rex, I. V. Gornyi, and A. D. Mirlin, {\it Majorana modes
in emergent-wire phases of helical and cycloidal magnet-superconductor
hybrids}, Phys. Rev. B {\bf 102}, 224501 (2020). 

\bibitem{Nothhelfer2022} J. Nothhelfer, S. A. D\'iaz, S. Kessler, T. Meng, M. Rizzi, K. M. D. Hals, and K. Everschor-Sitte, {\it Steering Majorana braiding via skyrmion-vortex pairs: A scalable platform}, Phys. Rev. B {\bf 105}, 224509 (2022).

\bibitem{Rokhinson2023} S. T. Konakanchi, J. I. V\"ayrynen, Y. P. Chen,
P. Upadhyaya, and L. P. Rokhinson, {\it Platform for
braiding majorana modes with magnetic skyrmions},
Phys. Rev. Res. {\bf 5}, 033109 (2023).

\bibitem{SpinWavesBook} A. I. Akhiezer, V. G. Bar'yakhtar, and S. V. Peletminskki, {\it Spin Waves}, John Wiley \& Sons (1968).

\bibitem{magnonics2015} S. A. Nikitov et al., {\it Magnonics: a new research area in spintronics and spin wave electronics}, Phys.-Usp. {\bf 58}, 1002 (2015). 

\bibitem{magnonics2021} A. Barman et al., {\it The 2021 magnonics roadmap}, J. Phys.: Condens. Matter {\bf 33}, 413001 (2021).

\bibitem{Brataas2020} A. Brataas, B. van Wees, O. Klein, G. de Loubens,
and M. Viret, {\it Spin insulatronics}, Phys. Rep. {\bf 885}, 1
(2020).

\bibitem{Nikitov2015} C. Davies, A. Francis, A. Sadovnikov, S. Chertopalov, M. Bryan, S. Grishin, D. Allwood, Y. Sharaevskii, S. Nikitov, and V. Kruglyak, {\it Towards graded-index magnonics: Steering spin waves in magnonic networks}, Phys. Rev. B {\bf 92}, 020408 (2015).

\bibitem{Braun1994} H.-B. Braun, {\it Fluctuations and instabilities of ferromagnetic domain-wall pairs in an external magnetic field}, Phys. Rev. B {\bf 50}, 16485 (1994).

\bibitem{Wulfhekel2004} R. Hertel, W. Wulfhekel, and J. Kirschner, {\it Domain-wall induced phase shifts in spin waves},
Phys. Rev. Lett. {\bf 93}, 257202 (2004).

\bibitem{Dijken2018} S. J. H\"am\"al\"ainen, M. Madami, H. Qin,  G. Gubbiotti, and  S. van Dijken, {\it Control of spin-wave transmission by a programmable domain wall}, Nat. Commun. {\bf 9}, 4853 (2018). 

\bibitem{Laliena2022} V. Laliena, A. Athanasopoulos, and J. Campo, {\it Scattering of spin waves by a Bloch domain wall: Effect of the dipolar interaction}, Phys. Rev. B {\bf 105}, 214429 (2022).

\bibitem{Nagaosa2014} J. Iwasaki, A. J. Beekman, and N. Nagaosa, {\it Theory of magnon-skyrmion scattering in chiral magnets}, Phys. Rev. B {\bf 89}, 064412 (2014).

\bibitem{Garst2014} C. Sch\"utte and M. Garst, {\it Magnon-skyrmion scattering in chiral magnets}, Phys. Rev. B {\bf 90}, 094423 (2014).

\bibitem{Aristov2015} D. N. Aristov, S. S. Kravchenko, A. O. Sorokin, {\it Magnon spectrum in ferromagnets with a skyrmion}, JETP Lett. {\bf 102}, 511 (2015).

\bibitem{Varma1998} T. K. Ng and C. M. Varma, {\it Spin and vortex dynamics and electromagnetic propagation in the spontaneous vortex phase}, Phys. Rev. B {\bf 58}, 11624 (1998).

\bibitem{Braude2004} V. Braude and E. B. Sonin, {\it Excitation of spin waves in superconducting ferromagnets},
Phys. Rev. Lett. {\bf 93}, 117001 (2004).

\bibitem{Golovchanskiy2018} I. A. Golovchanskiy, N. N. Abramov, V. S. Stolyarov, V. V. Bolginov, V. V. Ryazanov, A. A. Golubov, and A. V. Ustinov, {\it Ferromagnet/superconductor hybridization for
magnonic applications}, Adv. Func. Mater. {\bf 28}, 1802375 (2018).

\bibitem{Golovchanskiy2020} I. A. Golovchanskiy, N. N. Abramov, V. S. Stolyarov, V. V. Bolginov, V. V. Ryazanov, A. A. Golubov, and A. V. Ustinov, {\it Nonlinear spin waves in ferromagnetic/
superconductor hybrids}, J. Appl. Phys. {\bf 127}, 093903 (2020).

\bibitem{Golovchanskiy2020b} I. A. Golovchanskiy, N. N. Abramov, V. S. Stolyarov, V. I. Chichkov, M. Silaev, I. V. Shchetinin, A. A. Golubov, V. V. Ryazanov, A. V. Ustinov, and M. Yu. Kupriyanov, {\it Magnetization dynamics in proximity-coupled superconductor-ferromagnet-superconductor multilayers}, Phys. Rev. Applied {\bf 14}, 024086 (2020).

\bibitem{Bauer2022} T. Yu and Gerrit E. W. Bauer, {\it Efficient gating of magnons by proximity superconductors},
Phys. Rev. Lett. {\bf 129}, 117201 (2022).

\bibitem{Silaev2022} M. Silaev, {\it Anderson-Higgs mass of magnons in superconductor-ferromagnet-superconductor systems}
Phys. Rev. Applied {\bf 18}, L061004 (2022).
 
\bibitem{Golovchanskiy2023} I. A. Golovchanskiy, N. N. Abramov, O. V. Emelyanova, I. V. Shchetinin, V. V. Ryazanov, A. A. Golubov, and V. S. Stolyarov, {\it Magnetization dynamics in proximity-coupled superconductor-ferromagnet-superconductor multilayers. II. Thickness dependence of the superconducting torque}, Phys. Rev. Applied {\bf 19}, 034025 (2023)

\bibitem{Borst2023} M. Borst, P. H. Vree, A. Lowther, A. Teepe, S. Kurdi, I. Bertelli, B. G. Simon, Y. M. Blanter, 
and T. van der Sar, {\it Observation and control of hybrid spin-wave-Meissner-current transport modes}, 
Science {\bf 382}, 430 (2023).

\bibitem{Kharlan2024}  J. Kharlan, K. Sobucki, K. Szulc, S. Memarzadeh, and J. W. Klos, {\it Spin-wave confinement in a hybrid superconductor-ferrimagnet nanostructure}, Phys. Rev. Applied {\bf 21}, 064007 (2024).

\bibitem{Dobrovolskiy2019} O. V. Dobrovolskiy, R. Sachser, T. Br\"acher, T. B\"ottcher, V.
Kruglyak, R. V. Vovk, V. A. Shklovskij, M. Huth, B. Hillebrands,
and A. V. Chumak, {\it Magnon–fluxon interaction in a
ferromagnet/superconductor heterostructure}, Nat. Phys. {\bf 15},
477 (2019).

\bibitem{Berakdar2023} B. Niedzielski, C.L. Jia, and J. Berakdar, {\it Magnon-fluxon interaction in coupled superconductor/ferromagnet hybrid periodic structures}, Phys. Rev. Applied {\bf 19}, 024073 (2023).

\bibitem{Bobkova2022} I. V. Bobkova, A. M. Bobkov, A. Kamra, and W. Belzig, {\it Magnon-cooparons in magnet-superconductor hybrids}, Commun. Mater. {\bf 3}, 95 (2022).

\bibitem{Pearl1964} J. Pearl, {\it Current distribution in superconducting films carrying
quantized fluxoids}, Appl. Phys. Lett. {\bf 5}, 65 (1964).

\bibitem{Stern1967} F. Stern and W. E. Howard, {\it Properties of semiconductor surface inversion layers in the electric quantum limit}, Phys. Rev. {\bf 163}, 816 (1967).

\bibitem{Kuznetsov2023} M. A. Kuznetsov, K. R. Mukhamatchin, and A. A. Fraerman, {\it Effective interfacial Dzyaloshinskii-Moriya interaction and skyrmion stabilization in ferromagnet/paramagnet and ferro- magnet/superconductor hybrid systems}, Phys. Rev. B {\bf 107}, 184428 (2023).

\bibitem{AbrikosovBook} A. A. Abrikosov, {\it Fundamentals of the Theory of Metals} (North-Holland, Amsterdam, 1988).

\bibitem{Carneiro2000} G. Carneiro and E. H. Brandt, {\it Vortex lines in films: Fields and
interactions}, Phys. Rev. B {\bf 61}, 6370 (2000).

\bibitem{tanguy2001} C. Tanguy, {\it Counting the number of bound states of two-dimensional screened Coulomb potentials: A semiclassical approach}, arXiv:cond-mat/0106184.

\bibitem{Karnakov} V. Galitski, B. Karnakov, V. Kogan, V. Galitski Jr., {|it Exploring quantum mechanics: A collection of 700+ solved problems for students, lecturers, and researchers}, Oxford University Press 2013. 

\bibitem{Portnoi1997} M. E. Portnoi and I. Galbraith, {\it Variable-phase method and Levinson's theorem in two dimensions: Application to a screened Coulomb potential}, Sol. State Commun. {\bf 103}, 325 (1997).

\bibitem{Portnoi2005} D. G. W. Parfitt and M. E. Portnoi, {\it Exactly-solvable problems for two-dimensional excitons}, in Proceedings of the XI Regional Conference, Tehran, Iran, 3-6 May 2004: Mathematical Physics 52 (2005).

\bibitem{Makowski2011a} A. J. Makowski, {\it Bound states and quantization of screening in the Wannier-Mott excitons},
Phys. Rev. A {\bf 83}, 022104 (2011).

\bibitem{Makowski2011b} A. J. Makowski, {\it Quantum and classical solutions for statically screened two-dimensional Wannier-Mott excitons}, Phys. Rev. A {\bf 84}, 022108 (2011).
  
\bibitem{Lin1997} Q.-g. Lin, {\it Scattering by a Coulomb field in two dimensions}, Am. J. Phys. {\bf 65}, 1007 (1997).

\bibitem{Kotkin} G. L. Kotkin and V. G. Serbo, {\it Collections of problems in classical mechanics}, Pergamon Press 1971.

\end{thebibliography}
\end{document}